\documentclass[10pt]{article}

\usepackage{graphicx}
\usepackage{amsmath}
\usepackage{amssymb}
\usepackage{amsfonts}
\usepackage{epsfig}

\begin{document}

\author{\smallskip \sc T. A. S. Haddad\footnote{e-mail: 
{\tt thaddad@if.usp.br}} $^\dag$, S. T. R. Pinho$^\ddag$ and 
S. R. Salinas$^{\dag}$ \\ 
$^{\dag}$Instituto de F\'{\i}sica \\
Universidade de S\~{a}o Paulo \\
Caixa Postal 66318 \\
S\~{a}o Paulo - SP 05315-979, Brazil\\
$^\ddag$Instituto de F\'{\i}sica \\
Universidade Federal da Bahia\\
Salvador - BA 40210-340, Brazil}
\title{\bf Critical behavior of ferromagnetic spin models with aperiodic exchange
interactions}
\date{}
\maketitle

\begin{abstract}
We review recent investigations of the critical behavior of ferromagnetic 
$q$-state Potts models on a class of hierarchical lattices, with exchange
interactions according to some deterministic but aperiodic substitution
rules. The problem is formulated in terms of exact recursion relations on a
suitable parameter space. The analysis of the fixed points of these
equations leads to a criterion to gauge the relevance of the aperiodic
geometric fluctuations. For irrelevant fluctuations, the critical behavior
remains unchanged with respect to the underlying uniform models. In the
presence of relevant fluctuations, a non-trivial symmetric fixed point,
associated with the critical behavior of the uniform model, becomes fully
unstable, and there appears a two-cycle of the recursion relations. A
scaling analysis, supported by direct numerical thermodynamical
calculations, shows the existence of a novel critical universality class
associated with relevant geometric fluctuations.
\end{abstract}

The discovery of quasicrystals, and the design and investigation of magnetic
superlattices, provided strong motivation for the analysis of spin models
with aperiodic exchange interactions. Along the lines of the Harris
\cite{harris} criterion to gauge the influence of quenched disorder on the
critical behavior of simple ferromagnetic systems, Luck \cite{luck} has
proposed that sufficiently strong geometric fluctuations, associated with
deterministic but aperiodic interactions, may also change the ferromagnetic
critical behavior of the underlying uniform models. Luck's heuristic
criterion has indeed been checked and confirmed in a number of cases
(including the quantum Ising chain \cite{quantum}, and the two-dimensional
Ising model \cite{2dising}).

In a series of recent publications \cite{bjp,physa,jpa,pre,ijmp,cm}, we
considered ferromagnetic Ising and Potts models, with aperiodic exchange
interactions according to a variety of two-letter substitution rules, on
hierarchical Migdal-Kadanoff lattices \cite{griffiths}. Taking advantage of
the lattice structure, we were able to derive an exact expression for Luck's
criterion, and to analyze the novel critical behavior in the presence of
relevant geometrical fluctuations. We now present an overview of these
calculations.

As an example, consider the period-doubling two-letter substitution rule, 
\begin{equation}
A\rightarrow AB,  \label{pda}
\end{equation}
\begin{equation}
B\rightarrow AA,  \label{pdb}
\end{equation}
so that the successive application of this inflation rule on letter $A$
produces the sequence 
\begin{equation}
A\rightarrow AB\rightarrow ABAA\rightarrow ABAAABAB\rightarrow ....
\end{equation}
At each stage of this construction, the numbers $N_{A}$ and $N_{B}$, of
letters $A$ and $B$, can be obtained from the recursion relations 
\begin{equation}
\left( 
\begin{array}{c}
N_{A}^{\prime } \\ 
N_{B}^{\prime }
\end{array}
\right) =\mathbf{M}\left( 
\begin{array}{c}
N_{A} \\ 
N_{B}
\end{array}
\right) ,
\end{equation}
with the substitution matrix 
\begin{equation}
\mathbf{M}=\left( 
\begin{array}{cc}
1 & 2 \\ 
1 & 0
\end{array}
\right) .
\end{equation}
The eigenvalues of this matrix, $\lambda _{1}=2$, which is the period of the
transformation, and $\lambda _{2}=-1$, govern most of the geometrical
properties of the sequence. The total number of letters, at a large order $n$
of these constructions, behaves asymptotically as $\lambda _{1}^{n}$. The
geometric fluctuations are of the order $\left| \lambda _{2}\right| ^{n}$.
It is then convenient to define the wandering exponent of the sequence, 
\begin{equation}
\omega =\frac{\ln \left| \lambda _{2}\right| }{\ln \lambda _{1}},
\label{omega}
\end{equation}
that expresses the asymptotic dependence of the fluctuations with the total
number of letters, $\Delta N\sim N^{\omega }$.

The period-doubling sequence can be used to construct aperiodic spin models
on hierarchical diamond-type lattices with $b=\lambda _{1}=2$ bonds and $m$
branches. To go beyond the simple Ising model, and introduce an extra
parameter to work with, we consider a $q$-state ferromagnetic Potts model,
given by the Hamiltonian 
\begin{equation}
\mathit{H}=-q\sum_{(i,j)}J_{i,j}\delta _{\sigma _{i},\sigma _{j}},
\end{equation}
where $\sigma _{i}=1,2,...,q$, at all sites of the hierarchical lattice,
$J_{i,j}>0$, and the sum refers to nearest-neighbor pairs of sites
(for $q=2$, we recover the Ising model). The couplings can take only 
two values, $J_{A} $ and $J_{B}$, associated with a sequence of letters 
produced by the period-doubling substitution. In Fig. 1, we indicate some 
stages of the construction of this model on a simple diamond lattice 
(with $b=\lambda_{1}=2$ bonds and $m=2$ branches). Note that the choice of 
the same interactions along the branches of the diamonds is indeed mimicking 
an aperiodic layered structure.
\begin{figure}
\begin{center}
\epsfbox{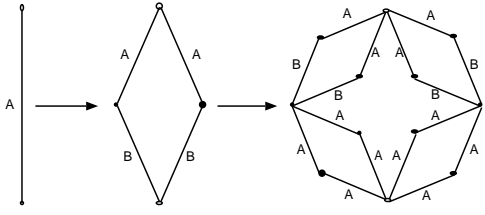}
\caption{Three successive generations of the simple diamond lattice ($b=m=2$).
Letters $A$ and $B$ indicate the exchange parameters (which are chosen
according to the period-doubling sequence).}
\end{center}
\end{figure}

We may also consider more general substitution rules and basic ``diamonds''
with $m$ branches and $b$ bonds along each branch, which leads to a lattice
with the intrinsic or fractal dimension 
\begin{equation}
D=\ln \left( mb\right) /\ln \left( b\right) .  \label{d}
\end{equation}
In each one of these structures, aperiodicity may be implemented by a
substitution rule of the form 
\begin{equation}
\left( A,B\right) \rightarrow \left(
A^{n_{1}}B^{b-n_{1}},A^{n_{2}}B^{b-n_{2}}\right) ,
\end{equation}
with $0\leq n_{1},n_{2}\leq b$. These sequences are characterized by a
substitution matrix with eigenvalues $\lambda _{1}=b$ and $\lambda
_{2}=n_{2}-n_{1}$, which leads to the wandering exponent 
\begin{equation}
\omega =\frac{\ln \left| \lambda _{2}\right| }{\ln \lambda _{1}}=\frac{\ln
\left| n_{2}-n_{1}\right| }{\ln b}.
\end{equation}

Now we decimate the internal degrees of freedom of the diamonds to write
exact recursion relations for the reduced couplings. For the $q$-state Potts
model on a lattice with $b=2$ bonds per branch, and with interactions
according to the period-doubling rule, given by Eqs. (\ref{pda}) and (\ref
{pdb}), it is easy to write \cite{pre} 
\begin{equation}
x_{A}^{\prime }=\left( \frac{x_{A}x_{B}+q-1}{x_{A}+x_{B}+q-2}\right) ^{m},
\end{equation}
and 
\begin{equation}
x_{B}^{\prime }=\left( \frac{x_{A}^{2}+q-1}{2x_{A}+q-2}\right) ^{m},
\end{equation}
where $x_{A,B}=\exp \left( q\beta J_{A,B}\right) $, with $\beta =1/k_{B}T$.

\begin{figure}
\begin{center}
\epsfbox{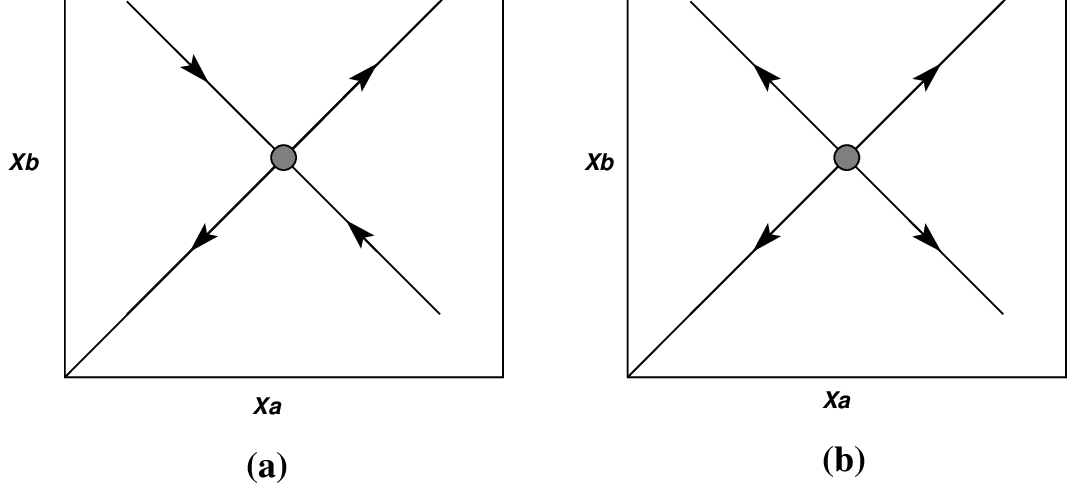}
\caption{Schematic representation of the parameter space and the flow
diagrams for irrelevant (a) and relevant (b) geometric fluctuations. The
non-trivial symmetric fixed point of saddle-point character for irrelevant
fluctuations (a) becomes fully unstable in the relevant case (b).}
\end{center}
\end{figure}
In parameter space, besides the trivial fixed points associated with zero
and infinite temperatures, there is a non-trivial symmetric fixed point,
that governs the critical behavior of the underlying uniform model (see the
sketch drawn in Fig. 2). The linearization of the recursion relations about
this symmetric fixed point leads to the matrix form 
\begin{equation}
\left( 
\begin{array}{c}
\Delta x_{A}^{\prime } \\ 
\Delta x_{B}^{\prime }
\end{array}
\right) =C\left( q\right) \mathbf{M}^{T}\left( 
\begin{array}{c}
\Delta x_{A} \\ 
\Delta x_{B}
\end{array}
\right) ,
\end{equation}
where $C\left( q\right) $ is a $q$-dependent structure factor, and 
$\mathbf{M}^{T}$ is the transpose of the substitution matrix. The eigenvalues
of this linear form may be written as 
\begin{equation}
\Lambda _{1}=C\left( q\right) \lambda _{1}=C\left( q\right) b,
\end{equation}
and
\begin{equation}
\Lambda _{2}=C\left( q\right) \lambda _{2}=C\left( q\right) \lambda
_{1}^{\omega }=C\left( q\right) b^{\omega },  \label{lambda2}
\end{equation}
where we have used the definition of $\omega $, given by Eq. (\ref{omega}).
For irrelevant geometric fluctuations, in which case the critical behavior
remains unchanged with respect to the underlying uniform model, it is
required that $\Lambda _{1}>1$, and $\left| \Lambda _{2}\right| <1$ (see
Fig. 2a). Therefore, we can write 
\begin{equation}
\Lambda _{1}=C\left( q\right) b=b^{y_{T}},  \label{lambda1}
\end{equation}
with the usual thermal critical exponent associated with the uniform system, 
\begin{equation}
y_{T}=\frac{D}{2-\alpha _{u}},  \label{yt}
\end{equation}
where $D$, given by Eq. (\ref{d}), is the fractal dimension of the lattice,
and $\alpha _{u}$ is the critical exponent associated with the divergence of
the specific heat of the uniform model \cite{melrose,derrida}.
From Eqs. (\ref{lambda1}) and (\ref{yt}), we have the structure factor 
\begin{equation}
C\left( q\right) =b^{\frac{D}{2-\alpha _{u}}-1}.
\end{equation}
Inserting this expression into Eq. (\ref{lambda2}) for the second
eigenvalue, we have 
\begin{equation}
\Lambda _{2}=b^{\frac{D}{2-\alpha _{u}}-1+\omega },
\end{equation}
which leads to an analog of Luck's criterion for a hierarchical lattice. As
we have already mentioned, fluctuations are irrelevant for $\left| \Lambda
_{2}\right| <1$. However, if $\left| \Lambda _{2}\right| >1$, which
corresponds to a fully unstable symmetric fixed point, geometric
fluctuations are relevant (see Fig. 2b). Therefore, the critical behavior
departs from the uniform case if we have 
\begin{equation}
\left| \Lambda _{2}\right| >1\longrightarrow 
\omega >1-\frac{D}{2-\alpha _{u}},
\end{equation}
which is an exact criterion of relevance for the Migdal-Kadanoff
hierarchical lattices. For the Potts model on the simple diamond lattice,
with $b=2$ bonds and $m=2$ branches, this criterion can be written as 
\begin{equation}
q>q_{c}=4+2\sqrt{2}=6.828427...,
\end{equation}
where $q_{c}$ is the same critical value for the relevance of quenched
disorder according to a calculation by Derrida and Gardner \cite{derridag} in
the weak disorder limit (also, note that $\alpha _{u}>0$ for a uniform Potts
model on a simple diamond lattice with $q>q_{c}$).

The recursion relations are so simple that it turns out to be relatively
easy to carry out an analysis of the critical behavior under relevant
geometric fluctuations. It is not difficult to show that, for $q>q_{c}$, in
which case the non-trivial symmetric fixed point is fully unstable, there is
a two-cycle that seems to be the candidate to define a novel class of
critical behavior \cite{pre}. In terms of the second iterates of the
recursion relations, of which each point of this two-cycle is a fixed point,
it displays a saddle-point character, with stable and unstable manifolds.
Let us use standard scaling arguments to analyze the novel critical
behavior. In the thermodynamic limit, the reduced free energy per bond can
be written in the scaling form \cite{derrida,dee} 
\begin{equation}
f\left( x\right) =g\left( x\right) +\frac{1}{b^{2D}}f\left( x^{\prime \prime
}\right) ,
\end{equation}
where $g\left( x\right) $ is a regular function, $x^{\prime \prime }$ is a
second iterate of the recursion relations, $b$ is the rescaling factor
(number of bonds of the diamond hierarchical lattice), $D$ is the fractal
dimension, and we use the factor $b^{2D}$ because we need two iterates to go
back to the neighborhood of the initial point in parameter space. This
equation has the asymptotic solution \cite{dee} 
\begin{equation}
f\left( x\right) \approx \left| x-x^{\ast }
\right| ^{2-\alpha }P\left( \frac{
\ln \left| x-x^{\ast }\right| }{\ln \Lambda _{cyc}}\right),
\label{freenergy}
\end{equation}
where $x^{\ast }$ is one of the points of the two-cycle, $\Lambda _{cyc}$ is
the largest eigenvalue of the linearization of the second iterate of the
recursion relations about any one of the points of the cycle, $P\left(
z\right) $ is an arbitrary function of period $1$, and the critical exponent 
$\alpha $, associated with the specific heat, is given by 
\begin{equation*}
\alpha =2-2\frac{\ln b^{D}}{\ln \Lambda _{cyc}}=2-2\frac{\ln \left(
mb\right) }{\ln \Lambda _{cyc}}.
\end{equation*}
All the calculations that we have performed, for a $q$-state Potts model on
a variety of Migdal-Kadanoff hierarchical lattices, with different values of 
$b$ and $m$, and suitable substitution sequences, point out that the values
of $\alpha $ from this scaling analysis are unequivocally different from the
values of $\alpha _{u}$ for the corresponding underlying uniform models 
\cite{pre}. For example, for the period-doubling rule and the simple diamond
lattice ($b=m=2$), with $q=7$, the two-cycle is located at 
$(x_{A},x_{B})_{1}=\left( 5.285...\text{, }7.642...\right) $, and 
$(x_{A},x_{B})_{2}=\left( 6.697...\text{, }4.750...\right) $, with
eigenvalues of the second iterate $\Lambda _{1}=3.933...$ and $\Lambda
_{2}=0.985...$, from which we have $\alpha =-0.0022...$, which should be
compared with $\alpha _{u}=0.010...$. With $q=25$, the weakening of the
second-order transition is even clearer: the two-cycle is located at 
$(x_{A},x_{B})_{1}=\left( 6.942...\text{, }234.34...\right) $, and 
$(x_{A},x_{B})_{2}=\left( 39.023...\text{, }3.831...\right) $, with
eigenvalues of the second iterate $\Lambda _{1}=4.243...$ and $\Lambda
_{2}=0.343...$, from which we have $\alpha =0.0817...$, to be compared with 
$\alpha _{u}=0.404...$.

To test the validity of the scaling arguments, and the role of the two-cycle
as the responsible for the new critical behavior, we have performed direct
numerical analyses of the singularity of the free energy \cite{pre}. In these
problems, it is well known that the reduced free energy can be expressed as
an infinite series \cite{derrida,dee}. For the Potts model on the $b=2$
diamond-type lattice, with the period-doubling rule, we have the series 
\begin{equation}
f\left( x_{A},x_{B}\right) =\sum_{n=0}^{\infty }\frac{1}{\left( 2m\right)
^{n}}\left[ \frac{1}{3}\ln \left( x_{A}^{\left( n\right) }+x_{B}^{\left(
n\right) }+q-2\right) +\frac{1}{6}\ln \left( 2x_{A}^{\left( n\right)
}+q-2\right) \right] ,
\end{equation}
where $x_{A,B}^{\left( n\right) }$ indicate the $n$th iterate of the
recursion relations. As usual, we assume uniform convergence, and
differentiate term by term to obtain the specific heat per bond. The
critical temperature can be determined with high precision by making use of
the existence of the paramagnetic fixed point at $T=0$, corresponding to 
$x_{A,B}=\infty $, which causes the apparent divergence of the series if
summed without using any regularization trick. Fixing the parameters $q$ and 
$m$, and also the strengths of $J_{A}$ and $J_{B}$, the critical temperature
thus calculated in fact locates $x_{A,B}$ in the attraction basin of the
two-cycle. According to the expectations, for irrelevant geometric
fluctuations, or for the uniform model, this method yields a critical
temperature that locates the system in the stable manifold of the uniform
fixed point.

The singularity in the specific heat can be analyzed by fitting the
numerical thermodynamic data to a function of the form $A+B\left| \left(
T-T_{c}\right) /T_{c}\right| ^{-\alpha }$ over a somewhat arbitrary scaling
region. For the irrelevant and uniform cases, we have always obtained very
good fittings, in excellent agreement with the values of $\alpha _{u}$
predicted by the scaling theory around the uniform fixed point. Of course,
these fitted values did not present any detectable sensitivity on the
particular values of $J_{A}$ and $J_{B}$. The Ising model ($q=2$) on the
simple $b=m=2$ diamond lattice, with exchange interactions according to the
period-doubling sequence, had been previously and independently analyzed by
Andrade \cite{andrade1}, with the same conclusions. For larger values of $q$,
in which cases the scaling analysis predicts positive values of $\alpha $
(although, of course, smaller than the values of $\alpha _{u}$), the
fittings presented excellent agreement with the scaling predictions. For
weaker singularities (mainly negative values of $\alpha $), the fitted
values were always somewhat bigger than the scaling predictions, with better
agreement for increasing values of $q$. For $b=m=2$, and $q=7$, we obtained 
$\alpha =-0.005(4)$, to be compared with the scaling result $\alpha
=-0.0022...$; for $q=25$, we obtained $\alpha =0.08(2)$, to be compared with 
$\alpha =0.0817...$. From the numerical calculations, we have seen an
oscillatory behavior of the specific heat as a function of $T$. The period
of oscillation is roughly given by the argument in Eq. (\ref{freenergy}),
with better agreement for increasing values of $q$. These log-periodic
oscillations, which are well known phenomena associated with hierarchical
structures, and indicate the need of a more refined scaling analysis, have
also been detected in the calculations of Andrade \cite{andrade1,andrade2}.

In conclusion, we have shown the existence of an exact criterion to gauge
the relevance of geometric fluctuations in a class of spin systems on
Migdal-Kadanoff hierarchical lattices. For irrelevant fluctuations, the
critical behavior remains unchanged with respect to the underlying uniform
models. In the presence of relevant fluctuations, the non-trivial symmetric
fixed point, which is associated with the critical behavior of the uniform
model, becomes fully unstable, and there appears a two-cycle of the
recursion relations in parameter space. A standard scaling analysis,
supported by direct thermodynamical calculations, shows the existence of a
novel universality class associated with the relevant geometric fluctuations.

This work has been supported by the Brazilian agencies FAPESP, CAPES, and
CNPq.

\end{document}